\documentclass[12pt,a4paper,final]{iopart}
\usepackage{stackrel}
\usepackage{iopams}
\usepackage{graphicx}
\usepackage{upgreek}
\usepackage[breaklinks=true,colorlinks=true,linkcolor=blue,urlcolor=blue,citecolor=blue]{hyperref}

\renewcommand{\vec}[1]{\boldsymbol{#1}}
\newcommand*\mean[1]{\overline{#1}}

\begin{document}

\title{Equations of state from individual one-dimensional Bose gases}

\author{F.~Salces-Carcoba$^1$, C.~J.~Billington$^{1,2}$, A.~Putra$^1$, Y.~Yue$^1$, S.~Sugawa$^{1}$\footnote{Present address: PRESTO, Japan Science and Technology Agency (JST), Saitama 332-0012, Japan, and Graduate School of Science, Kyoto University, Kyoto 606-8502, Japan}, I.~B. Spielman$^1$}
\address{$^1$Joint Quantum Institute, National Institute of Standards and Technology, and University of Maryland College Park, Gaithersburg, Maryland, 20899, USA}
\address{$^2$School of Physics and Astronomy, Monash University, Victoria 3800, Australia}

\begin{abstract}
	We trap individual 1D Bose gases and obtain the associated equation of state by combining calibrated confining potentials with \emph{in-situ} density profiles. Our observations agree well with the exact Yang--Yang 1D  thermodynamic solutions under the local density approximation. We find that our final 1D system undergoes inefficient evaporative cooling that decreases the absolute temperature, but monotonically reduces a degeneracy parameter.
\end{abstract}

\vspace{2pc}
\maketitle
\section{Introduction}\label{introduction}

Strongly interacting systems are ubiquitous in modern physics, from astrophysical objects such as neutron stars to the myriad of correlated electron systems in condensed matter. Experimental developments in ultracold atomic physics enable multiple avenues to explore interacting quantum matter, for example with optical lattices~\cite{Bloch2008}, Feshbach resonances~\cite{Chin2008} or mediated long-range interactions~\cite{Lahaye2009}. Furthermore, tailored potentials can reduce the effective dimensionality of cold atomic gases; for example, a 2D optical lattice can partition a 3D system into an array of 1D gases~\cite{Paredes2004, Tolra2004, Kinoshita2004}. Remarkably, the theory of a 1D Bose gas (1DBG) with contact repulsive interactions  is exactly solvable at all temperatures~\cite{Lieb1963, Yang1969}, making it an ideal system to benchmark experiment against theory.

In cold atomic gases, the ${\approx 5\, {\rm nm}}$ range~\cite{Chin2008} of the interatomic potential is vastly smaller than the ${\gtrsim 100\, {\rm nm}}$ interatomic spacing, hence interactions are well described by a local contact potential with strength ${g}$. This gives the 1D Hamiltonian
\begin{equation}
H = \sum_{i}  \left[-\frac{\hbar^2}{2m}\frac{\partial^2}{\partial z^{2}_{i}} + V(z_i)\right]
 + \frac{g}{2} \sum_{i \neq j} \delta (z_i - z_j),
\end{equation}
for $N$ interacting bosons of mass $m$ which in the absence of a potential, i.e. ${V(z)=0}$, becomes the Lieb--Liniger~\cite{Lieb1963} Hamiltonian. 

Lieb and Liniger showed that eigenstates of this Hamiltonian are parametrized by the dimensionless interaction parameter ${\gamma = mg/ \hbar^2n}$, where ${n}$ is the density. Here the relevance of interactions increases with decreasing density. For ${\gamma \ll 1}$ mean-field theory accurately describes the system, while for ${\gamma \gg 1}$ the atoms strongly avoid one another and behave much like weakly interacting fermions. C.~N.~Yang and C.~P.~Yang extended this solution to nonzero temperature~\cite{Yang1969} and cold atom experiments have validated the accuracy of the ``Yang--Yang'' thermodynamics~\cite{vanAmerongen2008, Vogler2013, Yang2017}. 

\begin{figure}[htb!]
	\begin{center}
		\includegraphics[width=0.7\linewidth]{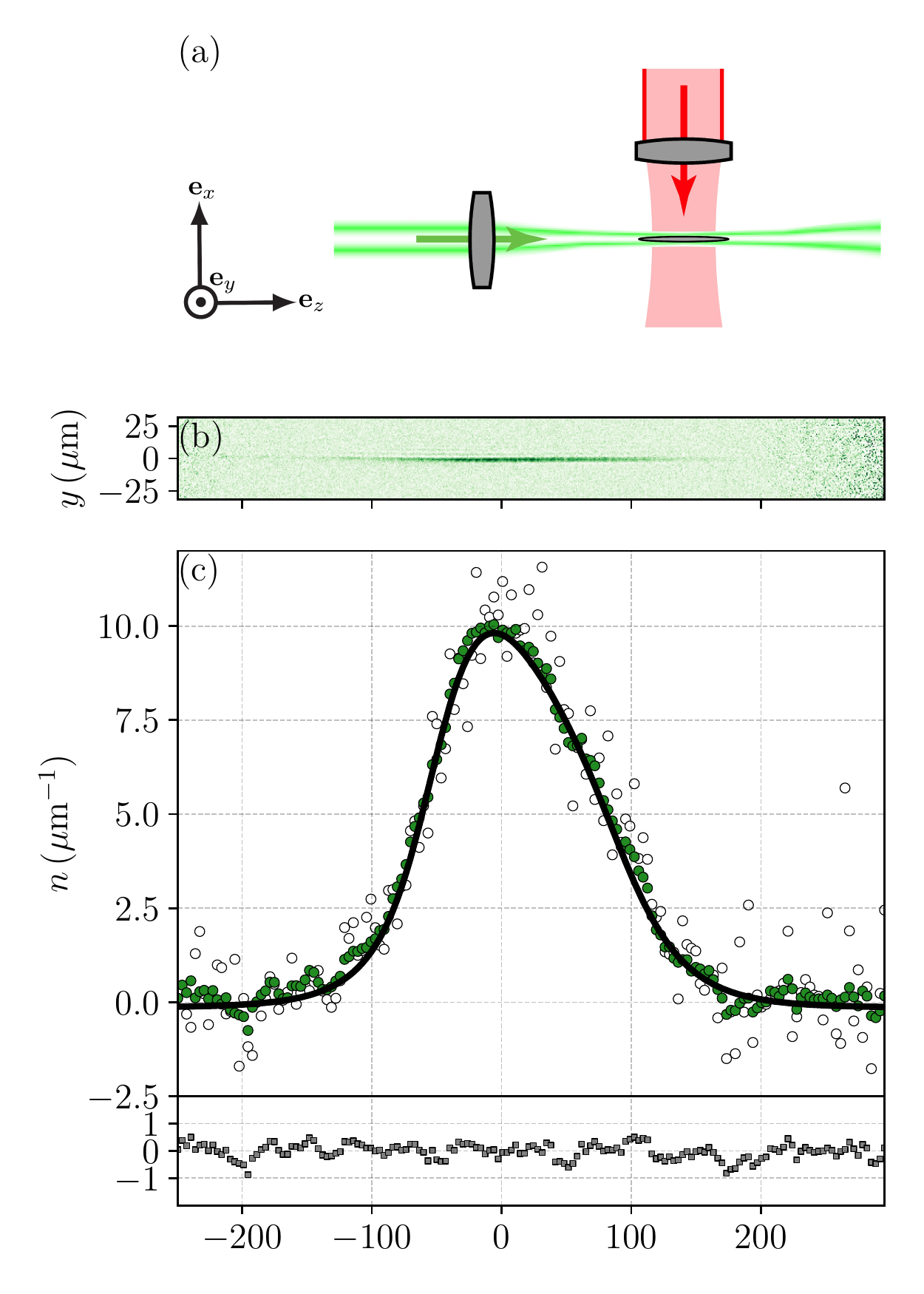}
		\caption[Fig. 1]{(a) Schematic of the trap formed by a tightly focused blue-detuned laser beam in the LG${_{01}}$ mode propagating along ${\bf e}_z$ and a red-detuned Gaussian laser beam propagating along ${\bf e}_x$. (b) \emph{In-situ} absorption image of the trapped 1DBG. (c) Linear densities from a single 1DBG (white circles),  the average of 100 realizations (green circles), and the fit of the Yang--Yang model to the average (black solid curve), along with the residuals (gray squares).}
		\label{fig:trap}
	\end{center}
\end{figure}

Here we study the physics of individual 1DBGs using $^{87}$Rb atoms in an optical ``tube trap'' (Figure~\ref{fig:trap}(a)) and benchmark the thermodynamic equation of state (EoS) against Yang--Yang thermodynamics. Our individual-system realization bridges an existing gap in experiments, on the one hand avoiding the issue of ensemble averaging present in realizations using optical lattices~\cite{Paredes2004, Tolra2004, Kinoshita2004, Vogler2013} and on the other hand enabling the future study of 1D multi-component systems not viable using magnetic confinement potentials~\cite{Rauer2016, Bongs2001}.

\begin{figure}[htb!]
	\begin{center}
		\includegraphics[clip, width=0.7\linewidth]{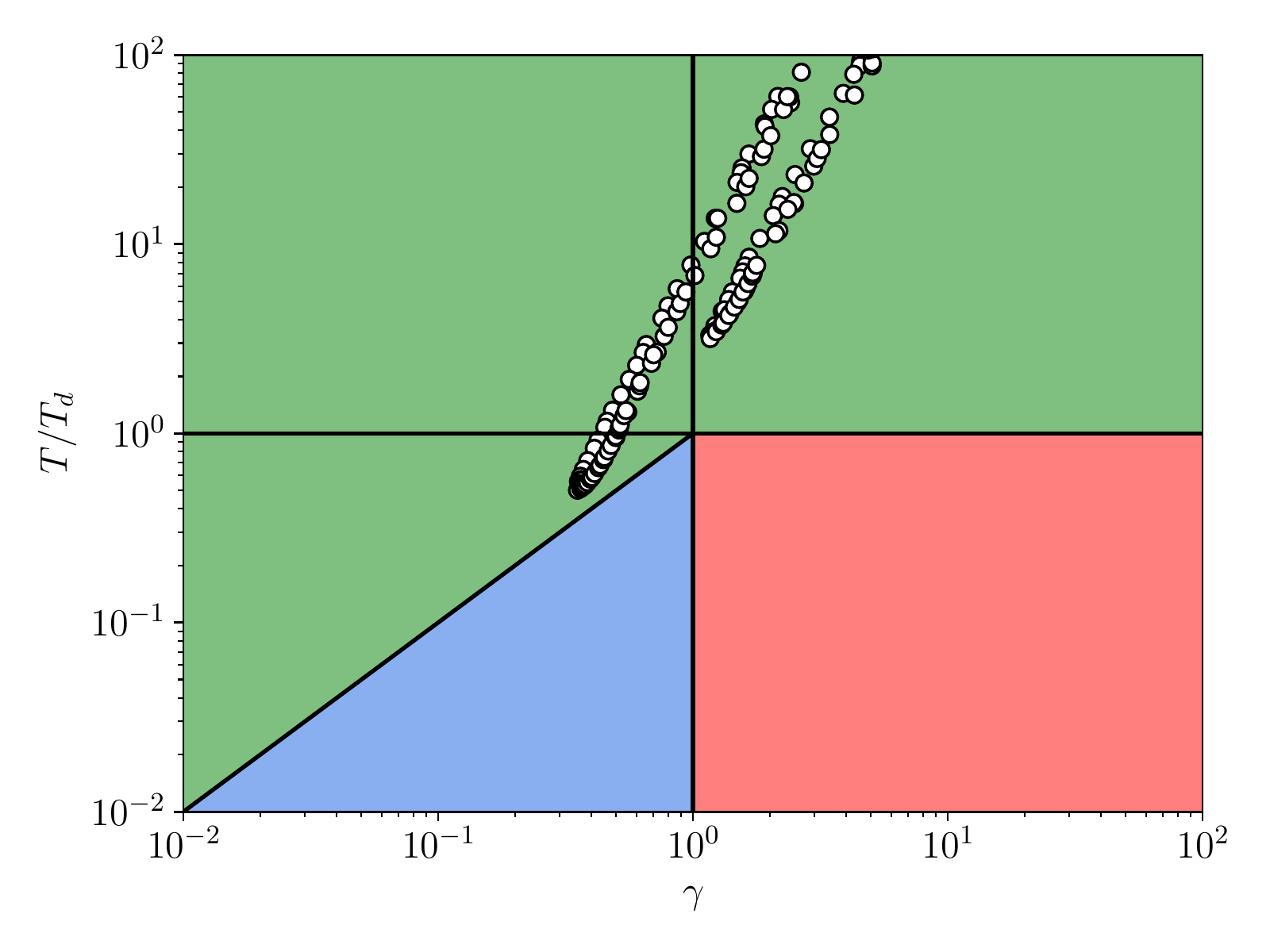}
	\end{center}
	\caption[Fig. 2]{Three regimes can be identified in the ${\gamma}$, ${T/T_d}$ parameter space which correspond to the strongly interacting degenerate (red), weakly interacting degenerate (blue) and non-degenerate (green) regimes. Specific experimental realizations of inhomogeneous systems cover a given range of these parameters (white circles).}	
	\label{fig:phases}
\end{figure}

The physics of 1DBGs can be divided into three qualitative regimes~\cite{Kheruntsyan2005} shown in Figure~\ref{fig:phases}. For sufficiently high temperatures (green region) the EoS is dominated by thermal effects and approaches that of a non-interacting classical gas. Below the degeneracy temperature ${T_d = \hbar^2 n^2 / 2m k_{\mathrm{B}}}$, where ${k_{\mathrm{B}}}$ is Boltzmann's constant, and for weak interactions (${\gamma \ll 1}$), the reduced thermal fluctuations allow Bose statistics to weigh in~\cite{Armijo2011}, creating a phase fluctuating degenerate gas. For lower temperatures where ${T/T_d < 2 \gamma}$ (blue region), the thermal energy falls below the chemical potential and the system is well described by the Gross--Pitaevskii equation (GPE).  In contrast, for systems with large interactions (${\gamma \gg 1 }$) the EoS resembles that of an ideal Fermi gas~\cite{Vogler2013}, with the formation of a Fermi surface for ${T < T_d}$ (red region). Yang--Yang thermodynamics provides EoS's encompassing all of these regimes, relating quantities like the particle, entropy and pressure densities to the chemical potential $\mu$ and temperature ${T}$, e.g., ${n(\mu, T)}$. 

In trapped systems, the confining potential ${V(z) \geq 0}$ can often be treated as an inhomogeneous chemical potential $\mu(z) = \mu_0 - V(z)$ which allows for multiple regimes to be present in a single 1DBG. We define ${\mu_0}$ as the local chemical potential at ${V(z) = 0}$. This can be quantitatively understood~\cite{Kohn1965} within the local density approximation (LDA) allowing the density profile $n(z, T)$ to be interpreted as an EoS $n(\mu(z), T)$. As a consequence, the EoS can be experimentally determined from a well-calibrated trapping potential $V(z)$ and the observed density profiles.

We extract this $n(\mu(z), T)$ EoS from \emph{in-situ} absorption images~\cite{Yefsah2011, Ku2012, Mukherjee2017} of individual systems (Figure~\ref{fig:trap}(b)), eliminating ensemble averaging. Because we obtain the 1D density directly, we do not apply the inverse Abel transform~\cite{Ku2012} thereby avoiding added noise. We benchmark our measurement against the Yang--Yang EoS (Figure~\ref{fig:trap}(c)), from which other thermodynamic quantities become readily available (e.g.~free energy, entropy and pressure).

This manuscript is organized as follows: in  Sec. (\ref{experiment}), we describe our experimental setup and data acquisition protocol; in Sec. (\ref{imaging}), we address the different calibration aspects of our analysis; in Sec.  (\ref{results}), we discuss the results; and in Sec. (\ref{conclusions}), we conclude.

\section{Experiment}\label{experiment}

We prepare cold atoms beginning with a magneto-optical trap followed by vertical magnetic transport~\cite{Greiner2001} into a magnetic quadrupole trap, ultimately loading a ${1064\ {\rm nm}}$ crossed optical dipole trap~\cite{YuJu2009, Starkey2013}. This gives ${N \approx 2 \times 10^5}$ atom Bose--Einstein condensates (BECs) of ${^{87}{\rm Rb}}$ in the ${5 \,\mathrm{S}_{1/2} \, |F=1, \, m_{F} = 0\rangle }$ hyperfine ground state with ${\approx 70\ {\rm Hz}}$ mean trapping frequencies. The atoms are then transferred into the composite high aspect-ratio trap shown in Figure~\ref{fig:trap}(a). This trap includes a red-detuned (${\lambda_{\rm G}=1064\ {\rm nm}}$) Gaussian beam along ${\bf e}_x$ with waist ${w_{\rm G} = 203(2)\ {\rm \upmu m}}$ providing reduced longitudinal confinement owing to its larger waist as compared to the ${\approx 70\ {\rm \upmu m}}$ crossed dipole beam waist. A transverse ``tube trap'' along ${\bf e}_z$ is provided by a blue-detuned (${\lambda_{\rm LG}=532\ {\rm nm}}$) Laguerre--Gauss (LG$_{01}$) beam, tightly focused to a waist of ${w_{\rm LG} = 5.6(5)\ {\rm \upmu m}}$. In our standard configuration these beams have powers ${P_{\rm G} = 0.8(1)\ {\rm W}}$ and ${P_{\rm LG} = 1.0(1)\ {\rm W}}$ , giving a peak transverse trapping frequency $\omega_{\perp} / 2\pi  =(\omega_{x}\omega_{y})^{1/2} / 2\pi = 17(2)\ {\rm kHz}$. 

\begin{figure}[hbt!]
	\begin{center}
		\includegraphics[width=0.7\linewidth]{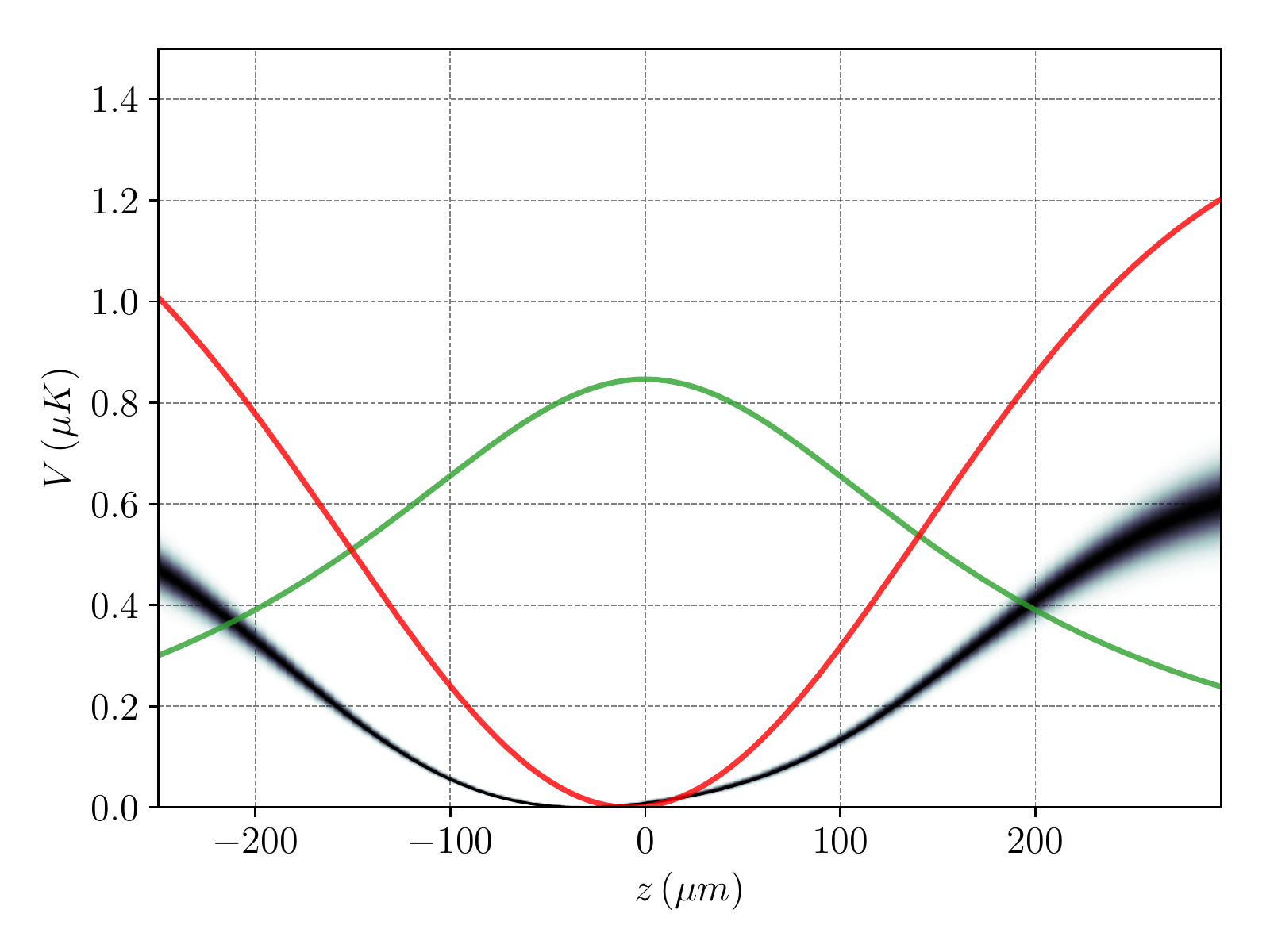}
		\caption[Fig. 4]{Trapping potential along ${{\bf e}_z}$ with both contributions from the anti-trap (solid green curve) and red-detuned beam (solid red curve). The shaded region marking the total trapping potential illustrates the uncertainty from the parameters entering into Eq. 2. This includes the covariance matrix for the parameters of ${V(z)}$ from our global Yang--Yang fit discussed in Sec.~(\ref{results}).}
		\label{fig:V}
	\end{center}
\end{figure}

The transverse zero-point energy from ${\omega_{\perp}}$ produces an anti-confining potential along ${\bf e}_z$ due to the divergence of the LG beam. The anti-trapping potential shown in green in Figure~\ref{fig:V} significantly alters the overall longitudinal potential
\begin{equation}
V(z) = \frac{\hbar\omega_{\perp}^{(0)}}{1+(z - z_a)^2/z_\mathrm{R}^2} + V_t \exp \left(-\frac{2z^2}{w_{\rm G}^2}\right) - V_0,
\end{equation} where $\omega_{\perp}^{(0)}/2\pi$ denotes the peak transverse trapping frequency; ${z_a}$ is the center of the anti-trap; ${z_\mathrm{R}}$ is the Rayleigh range of the LG beam; and ${V_0}$ is an energetic offset chosen such that the minimum of the potential is zero. The black shaded curve in Figure~\ref{fig:V} shows the combined anharmonic potential of the longitudinal trap (red curve) and the anti-confining potential. Small amplitude longitudinal dipole oscillations in the combined potential have frequency ${\omega_{z}/2\pi = 12.1(2)\ \mathrm{Hz}}$.

\begin{figure}[hbt!]
	\begin{center}
		\includegraphics[width=0.7\linewidth]{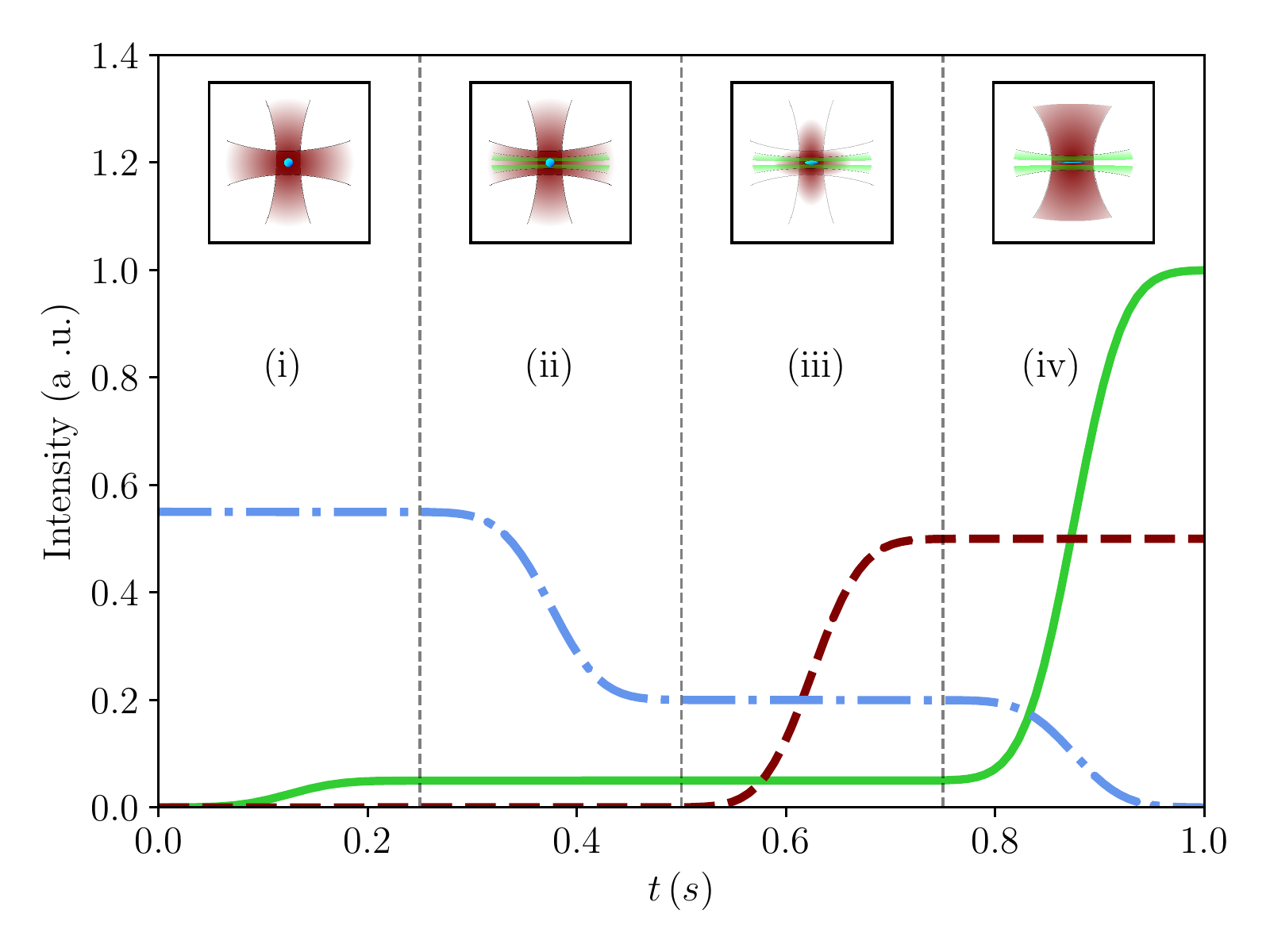}
		\caption[Fig. 3]{Adiabatic loading procedure. Each curve shows the intensity of a laser beam. The dash-dotted blue curve depicts the crossed dipole trap; the solid green curve denotes the LG tube trap; the dashed red curve marks the longitudinal trapping laser.}
		\label{fig:ramps}
	\end{center}
\end{figure}

Figure~\ref{fig:ramps} depicts our four step loading scheme. (i) We first ramp up the intensity of the LG beam from zero in ${250\ {\rm ms}}$ until the tube trap can suspend atoms against gravity. Because the 3D system is always larger than ${30 {\rm \upmu m}}$, the ${\approx 5\ {\rm \upmu m} }$ LG beam only captures a small fraction of the initial 3D ensemble. (ii) We then lower the intensity of the crossed dipole trap in ${250\ {\rm ms}}$, allowing the atoms outside the tube trap to fall away. (iii) We then ramp up the final ${1064\ {\rm nm}}$ longitudinal trap in ${250\ {\rm ms}}$. (iv) In the final ${250\ {\rm ms}}$ we simultaneously increase the intensity of the LG beam to its final value while removing the crossed dipole trap.

These ${250\ {\rm ms}}$ ramps were chosen to be adiabatic with respect to all the confining potentials. Monopole and dipole modes of the 1DBG can be induced by both beam misalignment and excessive ramp rates in this scheme. Our ramp times were chosen to mitigate these residual excitations.

We control the temperature of the 1DBG by varying the temperature ${T_{3\mathrm{D}}}$ of the initial 3D Bose gas. We tune ${T_{3\mathrm{D}}}$ by adjusting the depth of the crossed dipole trap, covering the range from ${T_{3\mathrm{D}}=34\ \mathrm{nK}}$ to ${320\ \mathrm{nK}}$, with an observed BEC transition at ${T_{3\mathrm{D}}^{\rm c} \approx 160\, \mathrm{nK}}$. We determine ${T_{3\mathrm{D}}}$ with time-of-flight (TOF) measurements. The number of atoms in the 1DBG increases with decreasing ${T_{3\mathrm{D}}}$ due to the increasing 3D density as ${T_{3\mathrm{D}}}$ falls. Gravitational sag is a complicating factor: as the crossed dipole trap decreases, the 3D ensemble lowers due to gravity, but the vertical alignment of the tube trap does not. We mitigate this effect by increasing the crossed dipole power after the final evaporation such that the crossed dipole potential is in a fixed vertical position prior to loading the tube trap.

\section{Imaging}\label{imaging}

We derive the density ${n(z)}$ from \textit{in-situ} absorption images. Our imaging system has a resolution of ${1.85(5)\ \mathrm{\upmu m}}$ and magnification that maps one ${5.6\ \mathrm{\upmu m}}$ sensor pixel to ${0.84(1) \ \mathrm{\upmu m}}$ in the object plane. In preparation for imaging, we apply a ${20\ {\rm \upmu s}}$ repump pulse to transfer the ${5\ \mathrm{S}_{1/2}}$ ${|F=1,\ m_{F}=0\rangle}$ atoms into the ${5\ \mathrm{S}_{1/2}}$ ${| F=2\rangle}$ hyperfine manifold. We then image~\cite{Genkina2016} with a circularly polarized ${\lambda_p = 780 \, {\rm nm}}$ probe beam resonant with the ${5\ \mathrm{S}_{1/2}}$ ${| F=2 \rangle}$ to ${5\ \mathrm{P}_{3/2}}$ ${| F=3\rangle}$ transition for ${20\ {\rm \upmu s}}$ at an average intensity of ${I=2.5 I_{\rm sat}}$, where ${I_{\rm sat}=1.67\ \mathrm{mW/cm}^2}$ is the saturation intensity of the resonant atomic transition. An image of the probe beam following absorption ${I_a}$, the probe without atoms present ${I_p}$, and a dark frame with no probe present ${I_d}$, are each recorded on a charge-coupled device (CCD) camera. From these images we obtain the absorbed fraction ${f =(I_p - I_a )/ (I_p - I_d)}$. For each set of experimental parameters we repeat the experiment for ${\approx 100}$ realizations. 

Our image analysis is a multiple step process. We first preprocess the raw images to correct for background artifacts and improve the signal-to-noise ratio (SNR) by a factor of ${\approx 10}$. We then extract the linear densities using an absorption model that includes a modest Lamb--Dicke suppression. As compared to a na\"ive model, ${n(z)}$ increases by as much as ${30 \%}$. This process leaves our qualitative results unchanged.

We reconstruct an optimal ${I_{p}^{\rm opt}}$ for each ${I_a}$ as a linear sum of  ${I_{p}}$ from all realizations by minimizing the sum squared difference between ${I_{p}^{\rm opt}}$ and ${I_a}$ away from the atoms~\cite{ockeloen_detection_2010, li_reduction_2007}. This reconstruction reduces fringing due to vibrational motion that occurred between acquiring ${I_{a}}$ and ${I_{p}}$, along with shot noise present on each ${I_{p}}$. We use similar techniques to remove a systematic difference in dark counts between ${I_{a}}$ and ${I_{p}}$, as well as to account for structured read-out noise. We then compute mean absorbed fractions ${\bar f}$ over each set of experimental parameters, and use uncentered principal component analysis to further suppress shot noise. From ${\bar f}$ and a detector calibrated~\cite{Chin2017, reinaudi_strong_2007, Yefsah2011} in units of ${I_{\rm sat}}$, we compute `na\"ive' optical depths using the standard solution to the Beer--Lambert (BL) law~\cite{reinaudi_strong_2007}, which we sum along columns to produce `na\"ive' linear densities.

We take into account the fact that the transverse extent of our atom cloud (for a tube trap the radial confinement gives an extent of ${\sqrt{\hbar/m \omega_{\perp}}\approx 110 \, {\rm nm}}$) is below the resolution of both our imaging system and the optical scattering length ${\sqrt{3 \lambda_p^2 /2\pi^2} \approx 300 \, {\rm nm}}$. We further incorporate the transverse diffusive motion that atoms undergo during the imaging pulse. Each of these effects violates the assumptions underlying the BL law. In comparison with the na\"ive BL law, our model for the density agrees at low density but deviates up to ${30 \%}$ at higher densities. This process is described in greater detail in the supplementary material.

\section{Results}\label{results}

The results of our image processing are 1D density profiles ${n^{(j)}(z)}$ confined in the same trapping potential but with one of 24 different initial conditions labeled by ${j}$. In the local density approximation we expect that these density profiles can result from Yang--Yang thermodynamics. For each ${j}$, both the temperature ${T^{(j)}}$ and the overall chemical potential ${\mu_0^{(j)}}$ are in principle unknown because of the lack of suitable reservoirs. As a result we obtain these quantities from fits to the Yang--Yang EoS and assess their validity in terms of the reduced chi-squared. 

For each $j$, the Yang--Yang EoS predicts the complete density profile as shown in Figure~\ref{fig:trap}(c) with just two free parameters $T^{(j)}$ and $\mu_0^{(j)}$. We constrain the fit to the trapping region between the local maxima of ${V(z)}$. The potential is parametrized by the common set of parameters shown in Table~\ref{tab1}. We include some of these as additional parameters in our fit shared between all $j$. In Table~\ref{tab1} we show the calibrations by other measurements along with their uncertainties; these are provided as initial guesses and bounds to the Yang--Yang fit. An additional uncalibrated parameter $\updelta z$ accounts for a tiny displacement of the 1DBG relative to the center of $V(z)$ for times following the loading protocol. The inclusion of $\updelta z$ leaves the main results unchanged and its value lies within the relative alignment uncertainty of the trap centers. Different combinations of fixed parameters have no qualitative effect on the results. The third column in Table~\ref{tab1} shows the potential parameters derived from the Yang--Yang fit. We evaluate the goodness-of-fit with the reduced chi-squared ${\chi^2_{\nu} \approx 1.5}$. Lastly, the residuals of the fit show systematic variations in $I_a$, which are reflected by the gray pixels in Figure~\ref{fig:trap}(c).

\Table{\label{tab1} Table summarizing the different parameters of $V(z)$. The calibrated values and their uncertainties were used as the central values and bounds for the fit.}

\br
Parameter&Calibrated value&Value from fit&Calibration method\\
\mr
$\omega_{\perp}/{2\pi}$  & ${17\, (4)\, {\rm kHz}}$ & ${17 \,(2) \,{\rm kHz}}$& Transverse expansion in TOF  \\ 
$z_a$  &$0 \,(10)\, {\rm \upmu m}$ & ${-7.670 \,(8)\, {\rm \upmu m}}$& Alignment precision  \\
$z_R$  & ${185\,(29) \,{\rm \upmu m}}$ & ${185\,(5)\, {\rm \upmu m}}$& Intensity profile of LG beam \\
${V_t/k_B}$ & ${-1.17\, (25)\, {\rm \upmu K}}$ & ${-1.37\, (6)\, {\rm \upmu K}}$& Intensity profile of Gaussian beam \\ 
$w_G$  &${203\, (2)\, {\rm \upmu m}}$ &  & Intensity profile of Gaussian beam \\ 
$\omega_z/ 2 \pi$  $V(z)$ & ${12.13\, (20)\, {\rm Hz}}$ & & Small amplitude dipole oscillations \\
$\updelta z$ && $ 8.19 \, (30)\, {\rm \upmu m}$ \\
\br

\end{tabular}
\end{indented}
\end{table}

\begin{figure}[hbt!]
	\begin{center}
		\includegraphics[width=0.45\linewidth]{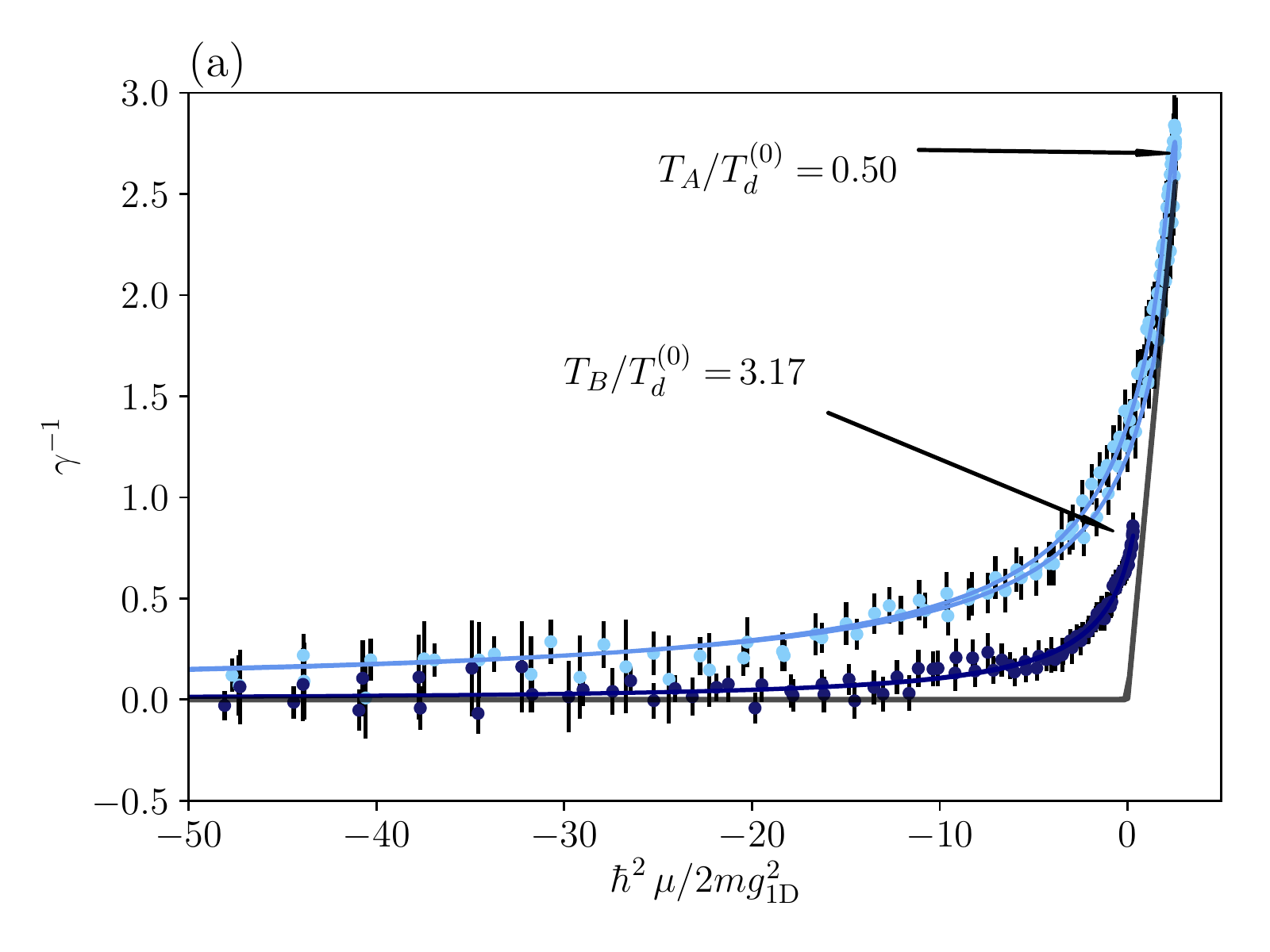}
		\includegraphics[width=0.45\linewidth]{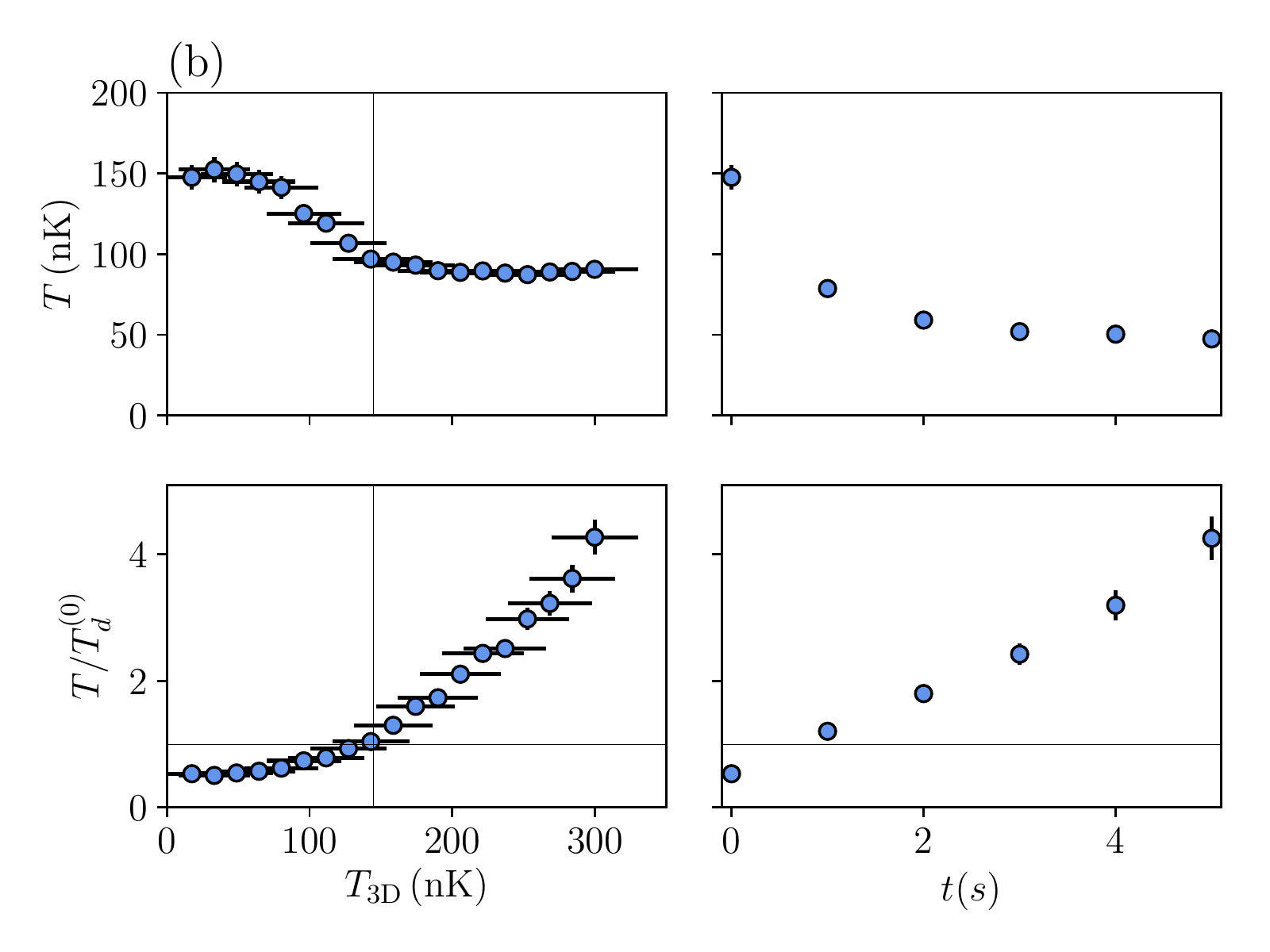}
		\includegraphics[width=0.45\linewidth]{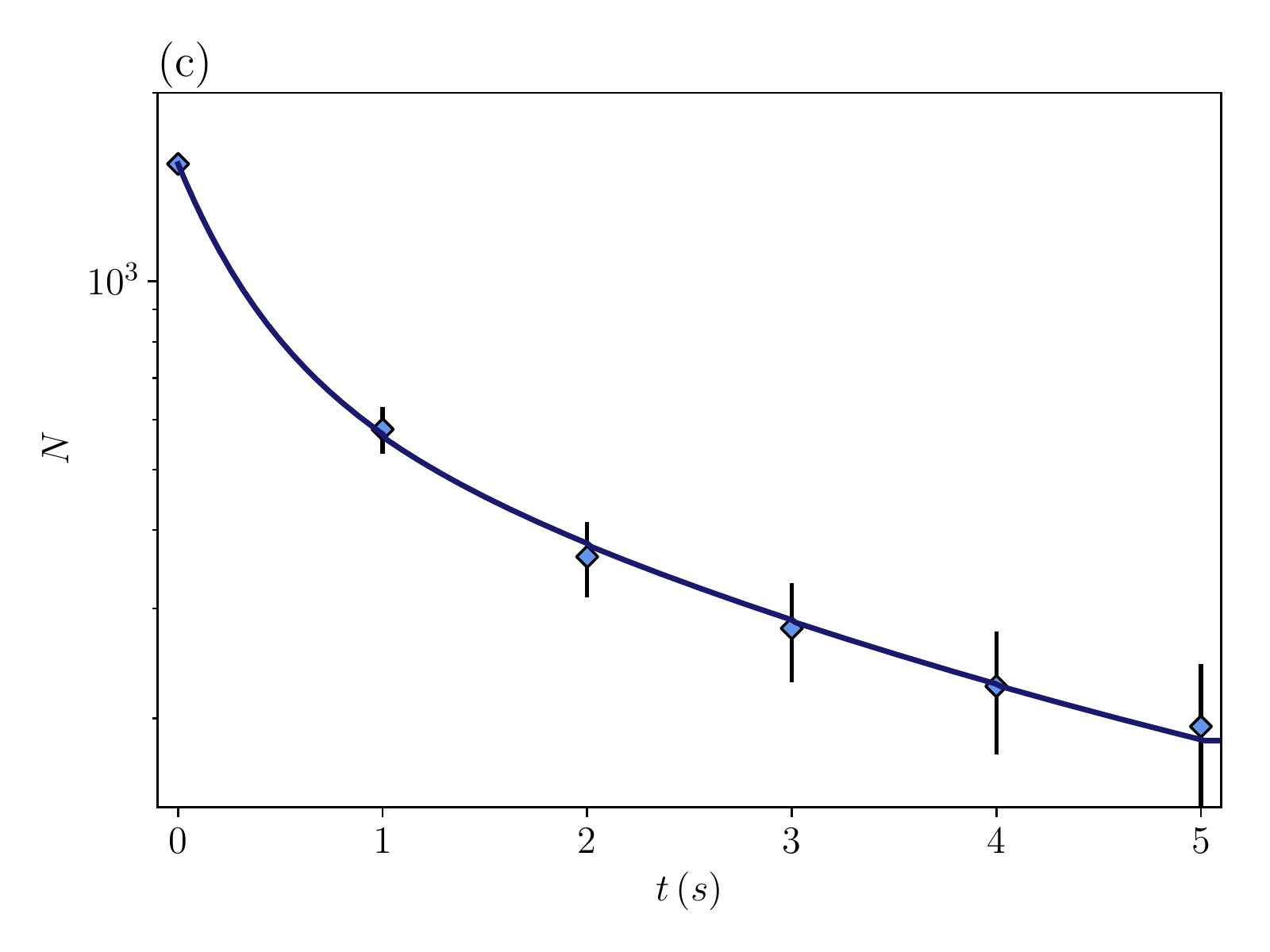}
		\caption[Fig. 5]{Results from Yang--Yang fit. (a) EoS for two different realizations (filled colored circles) along with Yang--Yang EoS (solid color curves) and mean-field prediction (solid gray curve). (b) Output parameters describing the state of the different ${T_{3\mathrm{D}}}$ realizations (left side panels) and the subsequent time evolution of the lowest ${T_{3\mathrm{D}}}$ realization (right side panels). The upper panels display the 1D absolute temperature ${T}$ while the bottom panels display the same temperature in units of the peak degeneracy temperature  ${T_d^{(0)}}$. (c) Total number of atoms as a function of time (blue diamonds) along with the fit to the one and three body loss model (solid dark-blue curve).}
		\label{fig:pars}
	\end{center}
\end{figure}

Figure~\ref{fig:pars}(a) shows the reduced density versus reduced chemical potential for two initial conditions, each plotting different cuts in the EoS $n(\mu, T)$. The continuous curves in Figure~\ref{fig:pars}(a) represent the Yang--Yang model with the $T$ and $\mu_0$ from our fits. For small chemical potential these density profiles are well described by the EoS of a non-interacting Bose gas while for $\mu > 0$ they approach the predictions of GPE mean-field theory~\cite{Kheruntsyan2005}. The Yang--Yang EoS accurately predicts both regimes. A sharp eye observes an apparent hysteresis loop visible in the trace labeled by $A$, this results from the spatial dependence of $g$ that follows $\omega_{\perp}$. As shown in Figure~\ref{fig:V}, $\omega_{\perp}$ is slightly off-centered, ultimately resulting in the observed behavior. The scattered white dots on Figure~\ref{fig:phases} represent these two traces in the $\gamma$, $T/T_d$ plane. These data are shown to be either in the interacting regime or below degeneracy, but not in both. 

Figure~\ref{fig:pars}(b) summarizes the outcome of all our Yang--Yang fits in which we varied $T_{3 \rm D}$, the initial 3D temperature (left panels); or the hold time $t$ in the 1D trap for our lowest $T_{3 \rm D}$ (right panels) cloud. In the top left panel, we observe that as a function of decreasing ${T_{3\mathrm{D}}}$, the 1D temperature $T$ first remains constant and then counterintuitively increases. In contrast, as shown in the bottom left panel, the degeneracy parameter defined as $T/T_d^{(0)}$, is a monotonically increasing function of $T_{3\mathrm{D}}$ showing how the more degenerate 3D clouds result in more degenerate 1DBGs.

For the lowest achievable $T_{3\mathrm{D}}$ and as a function of hold time, we see that both the total atom number $N$ and the 1D temperature $T$ drop (top right panel in Figure~\ref{fig:pars}(b) and Figure~\ref{fig:pars}(c) respectively), yet the 1DBG doesn't become more degenerate (bottom right panel in Figure~\ref{fig:pars}(b)). The simultaneous drop in $T$ and $N$ is consistent with evaporative cooling along the longitudinal axis of the tube-trap, which has depth of $\approx 700 {\rm nK}$. The inability of such evaporative cooling to increase or even maintain degeneracy results from the slower relative decrease in $T$ with respect to $T_d$ as driven by the atom number loss.

We explore the character of this loss by modeling the atom number decay with a model including one-body loss and three-body loss from photon scattering, background gas collisions and inelastic three-body collisions~\cite{Tolra2004}. 

Figure~\ref{fig:pars}(c) shows the measured atom number $N$ (blue diamonds). We fit the decay model to the observed number (dark blue curve), giving a one-body loss coefficient ${K_1^{1\mathrm{D}} = 0.108(2) s^{-1}}$ as well as a three-body loss coefficient ${K_3^{1\mathrm{D}} = 4.36(7) \times 10^{-29} \mathrm{cm}^{6} \mathrm{s}^{-1}}$. The value of $K_1^{1\mathrm{D}}$ is consistent with the combined vacuum-limited lifetime and estimated off-resonant scattering rate from the static dipole potentials. In contrast, the three-body loss coefficient from our fit is in excess of the intrinsic 3D three-body loss coefficient ${K_3^{3\mathrm{D}} = 5.8(3)\times 10^{-30} \mathrm{cm}^{6} \mathrm{s}^{-1}}$~\cite{Burt1997} by a factor of ${\approx 7.5}$. We attribute this enhancement to the difference in the three-body correlation function ${g^{(3)}}$~\cite{Tolra2004, Kheruntsyan2005} from a purely coherent sample. The observed cooling is consistent with initial rapid evaporation as atoms with sufficient kinetic energy~\cite{Fedichev1996} overcome the longitudinal barrier of $V(z)$. 

\section{Conclusions}\label{conclusions}
We realized individually trapped 1DBGs in a crossed dipole trap formed by a blue-detuned LG$_{01}$ beam and a red-detuned Gaussian beam. We benchmarked the EoS computed from Yang--Yang thermodynamics against the measured density profiles. We found that evaporative cooling along the edges of the tube trap took place although this did not maintain or increase the system's degeneracy. Our approach enables future exploration of spinor 1DBGs associated with multi-component physics~\cite{Li2015, Li2016}, including spin--orbit coupling~\cite{Cole2017}. This therefore presents a promising venue to study the limits of strongly interacting 1D systems in and out of equilibrium.

\section{ Acknowledgments}\label{acknowledgments}
We thank Max Schemmer for insightful discussions. We acknowledge the support for this work provided by the AROs Atomtronics MURI, NIST and the NSF through the PFC at the JQI. \\

\section*{\refname}
\bibliographystyle{unsrt}
\bibliography{bib_file}

\clearpage

\appendix 

\section{Image preprocessing and extraction of modeled linear densities}

We perform preprocessing of our raw images in order to improve signal to noise and correct for known systematic error, before extracting linear densities from each set of experimental parameters using an absorption model. The analysis pipeline from raw absorption, flat field and dark field images to linear densities is as follows. In the below descriptions of our analysis pipeline we alternately use bold symbols such as $\vec u$ when we are treating an image array as a vector for the purposes of linear algebra, and ordinary symbols with spatial dependence $u(x, y)$ when we are treating images as functions of space.

\subsection{Probe image reconstruction}

For each shot we reconstruct an optimal probe image $I_p^{\rm opt}$ as a linear sum of probe images from all shots:
\begin{equation}
\vec I_p^{\rm opt} = F\vec c,
\end{equation}
where $\vec I_p^{\rm opt}$ is $I_p^{\rm opt}$ unwrapped into a column vector, $F$ is a matrix containing all probe images as columns and $\vec c$ is a vector of coefficients. The optimal coefficients are determined by weighted linear regression:
\begin{equation}
(F^\mathsf{T}WF) \vec c = F^\mathsf{T}W\vec I_a,
\end{equation}
where $W$ is a diagonal matrix of weights equal to zero in a region of interest (ROI) about the atoms and one otherwise, and $\vec I_a$ is the image (as a column vector) from the shot in question of the probe with atoms present. The vector of coefficients $\vec c$ is determined by numerically solving the linear system, leading to an $I_p^{\rm opt}$ that minimizes the sum squared error with $I_a$ in the region outside the ROI. This probe reconstruction both reduces fringing due to vibrational motion that occurs between exposures within a shot, and reduces shot noise present on each reconstructed probe image on account of the dimensionality reduction entailed by linear regression~\cite{ockeloen_detection_2010, li_reduction_2007}.

\subsection{Dark field reconstruction}

We correct for a small spatially inhomogeneous systematic difference in counts ($\approx 1.2$ max, $\approx 0.2$ typical) between absorption and probe images, which we attribute to variation in ambient brightness over the 60Hz mains power cycle (this is systematic rather than random, as each shot is synchronized to begin at the same point in the 60Hz cycle). We fit a candidate two-dimensional function to the measured average difference between absorption and probe images, which we include as a reference image $\vec I_d^{\rm sys}$ in the above linear regression in order to extract a coefficient $c^{\rm sys}$ for each absorption image for how much of this offset was present (we obtain $c^{\rm sys} \approx 1$ in all cases indicating little shot-to-shot variation in the offset).

We then compute the mean dark frame $\mean{\vec I_d}$ over all shots, and perform principal component analysis (PCA) on the set of all dark frames, with two PCA eigenvectors revealing a source of correlated dark noise in the form of spatially sinusoidally varying dark counts with a different phase offset for each shot, which we also observed to be present in the PCA eigenvectors of the probe images (as eigenvectors four and five). We project each absorption image onto these eigenvectors $\vec I_p^{\rm pca4}$ and $\vec I_p^{\rm pca5}$ of the probe images in order to determine coefficients $c^{\rm pca4}$ and $c^{\rm pca5}$. A reconstructed dark field image $\vec I^{\rm recon}_d$ is then computed for each shot as
\begin{equation}
\vec I_d^{\rm recon} = \mean{\vec I_d}
+ c^{\rm sys} \vec I_d^{\rm sys}
+ c^{\rm pca4} \vec I_p^{\rm pca4}
+ c^{\rm pca5} \vec I_p^{\rm pca5}.
\end{equation}

\subsection*{Absorbed fraction and saturation parameter}
Absorbed fraction $f$ and saturation parameter $S$ images are then computed for each shot as
\begin{equation}
f = \frac{I_p^{\rm opt} - I_a}{I_p^{\rm opt} - I_d^{\rm recon}},
\end{equation}
\begin{equation}
S = \frac1{I_{\rm sat}}(I_p^{\rm opt} - I_d^{\rm recon}),
\end{equation}
where $I_{\rm sat}$ is the saturation intensity in camera count units. The `na\"ive' optical depth for a single shot can then be computed as:
\begin{equation}
{\rm OD}_{\rm na\ddot{\i}ve} = -\alpha \log\left(1 - f\right) + S f,
\end{equation}
where $\alpha \equiv \sigma_0/\sigma_{\rm eff} = I_{\rm sat\,eff}/I_{\rm sat}$ is the empirically determined ratio between the ideal two-level and effective scattering cross sections due to imperfect polarization and magnetic field orientation. The average na\"ive optical depth over all repeated shots for each set of experimental parameters is computed as: 
\begin{equation}
\mean{\rm OD}_{\rm na\ddot{\i}ve} \approx -\alpha \log\left(1 - \mean f\right) + \mean {SA},
\end{equation}
where the mean product of absorbed fraction and $\mean f$ and saturation parameter $\mean S$ are taken over only the repeated shots for one set of experiment parameters, and where we compute the mean absorbed fraction within the log rather than the mean of the entire log term in order to avoid the systematic error that results from taking the mean of a nonlinear function of noisy data. This na\"ive optical depth is not accurate across our entire dataset due to our 1D system being narrower transversely than both the optical scattering length $\sqrt{3\lambda_p^2/2\pi^2}$ and our imaging resolution, both of which are violations of the assumptions of the Beer--Lambert law. We continue to compute further reconstructions of this na\"ive optical depth only for comparison with modeled linear densities which include a correction to the Beer--Lambert law to account for this, presented further below.

\subsection{Dimensionality reduction of absorbed fraction}

We dimensionally reduce the mean absorbed fractions $\mean f$ of each set of experiment parameters in order to reduce the effect of shot noise on column sums of the data. Since the point spread function resulting from diffraction in our imaging system is fixed from shot to shot, this has the effect of projecting the measured absorbed fractions onto the empirically observed point spread function and its most common variations, suppressing spurious apparent absorption due to shot noise in regions where the point spread function results in little absorption.

The dimensionality reduction proceeds as follows. First we crop each mean absorption image $\mean f(x, y)$ to the $75$-pixel high region of interest that entirely contains our imaging system's point spread function to form $\mean f_{\rm ROI}(x, y)$. Then, for each $x$ position $x_i$ in the image, we extract the vertical lines of $\mean f_{\rm ROI}(x, y)$ from all sets of experiment parameters, at that $x$ pixel and the nearest four $x$ pixels. Treating each vertical line as a vector\footnote{The mixed vector vs. function-of-space notation here is because we are treating the vertical slices of the images as vectors, performing dimensionality reduction on a slice-by-slice basis, whereas the $x$ coordinate is merely a label selecting which vector we are referring to.} $\mean{\vec f}_{\rm ROI}(x_i) = \mean f_{\rm ROI}(x_i, y)$, we obtain the set of vectors $\{\mean {\vec f}_{\rm ROI}(x_{i+j}), -2 \leq j \leq 2\}$ and perform unentered PCA~\cite{cadima_relationships_2009}, keeping only the first four normalized eigenvectors $\{\hat {\vec v}_n(x_i), n \in [1, 4]\}$. We then project the (also cropped to the ROI) vertical lines $\vec f_{\rm ROI}(x_i)$ of the absorbed fractions for each individual shot at the original $x$ position onto the subspace spanned by these vectors:
\begin{equation}
f_{\rm red}(x_i, y) = {\vec f}_{\rm red}(x_i) = \sum_{n=1}^4 \left[\hat{\vec v}_n (x_i)\cdot {\vec f_{\rm ROI}}(x_i)\right]\hat{\vec v}_n (x_i).
\end{equation}
Thus we dimensionally reduce each vertical slice (within the ROI) of each shot's absorbed fraction onto a basis of four basis functions chosen by unentered PCA of the vertical slices of all mean absorbed fractions at that $x$ pixel and the nearest four other $x$ pixels. We then compute mean, dimensionally reduced absorbed fractions $\mean f_{\rm red}(x, y)$ for each set of experiment parameters within the ROI (Hereafter any images mentioned should be assumed to be cropped to the ROI).

We observe that to within numerical rounding error, it makes no difference whether the mean absorbed fractions are dimensionally reduced into this space, or the individual absorbed fractions are, before being averaged together again. We do the latter in order to provide a statistical uncertainty estimate in the mean absorbed fraction of a given set of experiment parameters as
\begin{equation}\label{eq:uncertainty_estimate_A}
\Delta\mean f_{\rm red}(x, y)
= \frac{\sigma\left[\{f_{\rm red}(x, y)\}\right]}{\sqrt{N}},
\end{equation}
where $\sigma$ is the standard deviation over all repeated shots for one set of experiment parameters, $\{A_{\rm red}(x, y)\}$ is the set of all dimensionally reduced absorbed fractions for those shots, and $N$ is the number of repeated shots for that set of experiment parameters.

\subsection*{Saturation parameter at the position of the atoms}
Due to the point spread function of our imaging system being larger than the vertical extent of our atomic cloud, the saturation parameter at the location of apparent absorption (after diffraction) does not correspond to the saturation parameter at the actual location of the atoms, which is where saturation effects are relevant. We estimate a saturation parameter $S_0(x)$ for each set of experiment parameters at the estimated $y$ position of the atoms by interpolating the mean saturation parameter for that set of experiment parameters to the $y$ position where there is maximum apparent absorption over all shots. The $y$ position of maximum apparent absorption at each $x$ position is taken to be a quadratic fit $y_0(x) = ax^2 + bx + c$ with parameters determined by maximizing the total absorption over all shots:
\begin{equation}
[a, b, c] = \stackrel[a, b, c]{}{\mathrm{argmax}}
\left(\sum_{\rm shots}\sum_{x_i}
\stackrel[y]{}{\mathrm {interp}}[S(x_i, y)](ax_i^2 + bx_i + c)
\right),
\end{equation}
where $\stackrel[y]{}{\mathrm {interp}}$ is a one-dimensional spline interpolation function interpolating in the $y$ direction only. The estimated saturation parameter at each $x$ position for each set of experiment parameters is then
\begin{equation}
S_0(x) = \stackrel[y]{}{\mathrm {interp}}[\mean S(x, y)](y_0(x)).
\end{equation}

\subsection{Na\"ive linear density}
We can now compute an improved na\"ive optical depth for each set of experiment parameters using the dimensionally reduced absorbed fractions and interpolated saturation parameter as:
\begin{equation}\label{eq:naive_reduced_od}
\mean{\rm OD}_{\rm na\ddot{\i}ve\,red}(x, y)
\approx -\alpha \log\left(1 - \mean f_{\rm red}(x, y)\right) + S_0(x) \mean f_{\rm red}(x, y),
\end{equation}
and then compute a na\"ive linear density $n_{\rm na\ddot{\i}ve}(x)$ at each $x$ position by dividing by the cross section and integrating along $y$:
\begin{equation}\label{eq:naive_linear_density}
n_{\rm na\ddot{\i}ve}(x) = \frac{\Delta y}{\sigma_0}\sum_{y} \mean{\rm OD}_{\rm naive\,red}(x, y),
\end{equation}
where $\Delta y$ is the pixel size.

% We can also compute an uncertainty estimate by propagating (\ref{eq:uncertainty_estimate_A}) through (\ref{eq:naive_reduced_od}) using linear uncertainty estimation and through (\ref{eq:naive_linear_density}) by adding in quadrature:
% \begin{equation}
% \Delta n_{\rm 1D\,na\ddot{\i}ve}(x) =
% \frac{\Delta y}{\sigma_0}\left(\sum_y\left(
% \frac{\alpha}{1 - \mean f_{\rm red}(x, y)} + S_0(x)\right)
% \Delta\mean f_{\rm red}(x, y)
% \right)^{\frac12}.
% \end{equation}

\subsection{Modeled linear density}
We face three related problems in computing the column density $n_{\rm col}(x, y)$ given a measured absorbed fraction $f(x, y)$ and saturation parameter $S(x, y)$ via the solution to the Beer--Lambert law~\cite{reinaudi_strong_2007}:
\begin{equation}
\sigma_0 n_{\rm col}(x, y) = -\alpha\log(1 - A(x, y)) + S(x, y) A(x, y)
\end{equation}
The first problem is that we do not measure $f(x, y)$ directly---we measure it only after it has diffracted in the $y$ direction, a difference which is not negligible given the size of our atom cloud in that direction. The second problem is that atoms do not only absorb light at their exact location in space, rather they absorb it from a surrounding region of space with cross sectional area given by the absorption cross section $\sigma_0$~\cite{striebel_absorption_2017}. The final problem is that our cloud is so small in the $y$ direction that diffusion of atoms during imaging may not be negligible. These latter two problems mean that we cannot infer $n_{\rm col}(x,y)$ from the usual solution to the Beer--Lambert law, we can only determine the convolution of $n_{\rm col}(x,y)$ with some absorption profile $g(y)$ that takes into account both the finite absorption region and the diffusion of atoms from their initial positions in the $y$ direction, the direction in which $g(y)$ is not small compared to our atom cloud. With this in mind, the solution to the Beer--Lambert law can be modified to read
\begin{equation}\label{eq:modified_beer_lambert}
\sigma_0 (n_{\rm col} * g)(x, y, t) = -\alpha\log(1 - f(x, y, t)) + S(x, y) f(x, y, t),
\end{equation}
where the convolution is only along the $y$ direction. Atomic diffusion and diffraction imply that the only quantity we have experimental access to is
\begin{equation}\label{eq:A_meas}
f_{\rm meas}(x) = \frac1\tau\int_0^\tau{\rm d}t\int{\rm d}y\, f(x, y, t)
= \sum_y
\mean f_{\rm red}(x, y)
\Delta y,
\end{equation}
that is, we only observe a time average of absorption over the imaging pulse time $\tau$, and we only observe the integral of the undiffracted absorbed fraction, since diffraction preserves this integral.

If the second term of the solution to the Beer--Lambert law dominates, then the na\"ive linear density is accurate, since all three of diffraction, diffusion and convolution preserve integrals of the absorbed fraction. It is only the log term that causes a problem, since its integral is not conserved under diffraction. 

Given a model for $(n_{\rm col} * g)(x, y, t)$ with a single parameter $n(x)$ for the linear density at each $x$ position and a measurement $f_{\rm meas}(x)$ for each $x$ position, we can invert (\ref{eq:A_meas}) and (\ref{eq:modified_beer_lambert}) to obtain the linear density, under the assumptions of the model.

Our model is the following: The absorption profile $g(y)$ is approximated by a Gaussian with unit integral and standard deviation $\sigma_y(t)$ equal at $t=0$ to the optical scattering length $\sqrt{\sigma_0/\pi} = \sqrt{3\lambda_p^2/2\pi^2}$ and increasing due to atomic diffusion as time elapses. Since the atom cloud is narrower than this absorption profile, we approximate the convolution $(n * g)(x, y, t)$ as:
\begin{equation}\label{eq:n_model}
(n * g)(x, y, t) \approx \frac{n_{\rm 1D}(x)}{\sqrt{2\pi\sigma_y^2(x, t)}}
\exp\left[-\frac{y^2}{2\sigma_y^2(x, t)}\right]
\end{equation}
where $\sigma_y^2(x, t)$ is increasing due to momentum diffusion:
\begin{equation}
\sigma_y^2(x, t) = \frac{\sigma_0}\pi + \frac13 \sigma_{v_y}^2(x, t) t^2,
\end{equation}
where the mean squared $y$ velocity $\sigma^2_{v_y}$ is given by the scattering rate $R_{\rm scat}$ and recoil velocity $v_{\rm rec}$:
\begin{equation}
\sigma^2_{v_y}(x, t) = \frac13(2\pi)^{-1}R_{\rm scat}(x) v_{\rm rec}^2 t,
\end{equation}
which is approximating isotropic scattering such that the per-scattering-event expected squared change in velocity is $v^2_{\rm rec}/3$.
The scattering rate, ignoring Doppler shifts away from resonance, is, in terms of the saturation parameter $S$:
\begin{equation}
R_{\rm scat}(x) = \frac\Gamma2\frac{S(x)}{1 + S(x)}.
\end{equation}
Putting this all together, the modeled $y$ variance of the absorption profile is:
\begin{equation}\label{eq:y_rms_model}
\sigma_y^2(x, t) = \frac{\sigma_0}{\pi} + \frac\Gamma{36\pi}\frac{S(x)}{1 + S(x)}v_{\rm rec}^2t^3.
\end{equation}
Over the duration of our imaging pulse, the diffusion described by the second term results in an increase in the absorption profile's standard deviation by $\approx 30\%$ compared to the effect of the non-zero optical scattering length alone.

Using our absorption model (\ref{eq:n_model}) with an absorption profile with $y$ variance given by (\ref{eq:y_rms_model}), and saturation parameter $S(x, y) $ given by our estimate $S_0(x)$, our modified Beer--Lambert law solution (\ref{eq:modified_beer_lambert}) becomes:
\begin{equation}
\sigma_0 n(x)\frac{\exp\left[-\frac{y^2}{2\sigma_y^2(x, t)}\right]}{\sqrt{2\pi\sigma_y^2(x, t)}}
 = -\alpha\log(1 - f(x, y, t)) + S_0(x) f(x, y, t),
\end{equation}
which, if numerically inverted, defines a function that takes only $n(x)$ as input and returns $f(x, y, t)$ at any given time. Numerically integrating the result in $t$ and $y$ as per (\ref{eq:A_meas}) extends this function into one which takes only $n(x)$ and returns the expected $f_{\rm meas}(x)$ for that linear density. Numerically inverting \emph{this} function then yields our final aim, of a function that takes as input $f_{\rm meas}(x)$ from our data and outputs a value of $n(x)$ for the linear density implied by the measured data and the model.

We perform the above computationally nontrivial calculation to extract modeled linear densities from our dimensionally-reduced mean absorbed fractions and interpolated saturation parameters for each set of experiment parameters.

The na\"ive and modeled linear densities agree at low densities but disagree by up to 20 percent at higher densities, with the na\"ive method underestimating linear densities compared to those obtained using the absorption model.

\section{Yang--Yang thermodynamics}

We use the Yang--Yang model~\cite{Yang1969} to describe our data. The exact diagonalization of the underlying Hamiltonian is carried out with the use of the thermodynamic Bethe ansatz (TBA) ($T > 0$ Bethe ansatz). From the TBA the following set of first-order integral equations can be derived
\begin{equation}
\epsilon(k) = \frac{\hbar^2k^2}{2m} - \mu - \frac{k_{\rm B} T}{2\pi} \int_{-\infty}^{\infty} \frac{2c}{c^2 + (k-q)^2} \, \ln(1+{\rm e}^{-\epsilon(q)/k_{\rm B} T}) \,{\rm d}q \, ,
\end{equation}
\begin{equation}
2\pi f(k)(1+{\rm e}^{\epsilon(k)/k_{\rm B} T}) = 1 + \int_{-\infty}^{\infty} \frac{2c}{c^2 + (k-q)^2} \,f(q) \, {\rm d}q \, ,
\end{equation}
\begin{equation}\label{eq:linear_density_YY}
n = \int_{-\infty}^{\infty} f(q)\, {\rm d}q \, ,
\end{equation}
where $m$ is the mass, $k_{\rm B}$ is the Boltzmann constant, $T$ is the temperature, $\mu$ the chemical potential and $c = m g/\hbar^2$ is the interaction wavenumber. Both $k$ and $q$ label momenta. The above equations can be solved recursively to compute $n$, the linear density, given the values for $\mu, T$ and $g$, the chemical potential, temperature and coupling constant. 

We implement a numerical solver for the YY equations within the LDA that takes the parameters $\mu$, $T$ as its primary input and computes $c$ and $g$ by using the appropriate values of the transverse trapping frequency $f_{\perp}$, the 3D scattering length $a_{\rm 3D}$ and mass $m$. We recursively solve for $\epsilon(k)$ and $f(k)$ from which we ultimately compute the density $n$. We transform all the momentum and energy quantities
\begin{equation}
\tilde{k} = k/\sqrt{2mk_{\rm B}T/\hbar^2},
\end{equation}
\begin{equation}
\tilde{E} = E/k_{\rm B} T,
\end{equation}
so that the first two YY equations read
\begin{equation}
\tilde{\epsilon}(\tilde{k}) = \tilde{k}^2 - \tilde{\mu} - \int_{-\infty}^{\infty} \frac{1}{\pi} \frac{\tilde{c}}{\tilde{c}^2 + (\tilde{k}-\tilde{q})^2} \, \ln(1+{\rm e}^{-\tilde{\epsilon}(\tilde{q})}) \,{\rm d}\tilde{q},
\end{equation}
\begin{equation}
2\pi f(\tilde{k})(1+{\rm e}^{\tilde{\epsilon}(\tilde{k})}) = 1 + \int_{-\infty}^{\infty} \frac{1}{\pi} \frac{\tilde{c}}{\tilde{c}^2 + (\tilde{k}-\tilde{q})^2} \,f(\tilde{q}) \, {\rm d}\tilde{q}.
\end{equation}

We denote the Lieb--Liniger kernel (a normalized Lorentzian) as $L(c, q)$. Our numerical solver performs a $k$-space convolution using the \texttt{scipy.signal.fftconvolve} method to evaluate the integrals. For this we use a $N_k = 1024$ points grid that covers the range $k = [-10\,k_{\rm th}, 10\,k_{\rm th}]$, where $k_{\rm th} = \sqrt{2mk_{\rm B}T/\hbar^2}$ is the thermal wavenumber. We initialize $\epsilon_0(k) = k^2 - \mu$ and iterate over the following recursive relation
\begin{equation}
\epsilon_{j+1}(k) = \epsilon_0(k) - L(c_0, q) \circledast \ln(1+{\rm e}^{-\epsilon_{j}(q)}),
\end{equation}
where $\circledast$ denotes the Fourier convolution operator. Once the convergence condition $ \sqrt{\sum_i(\epsilon_{i, j+1}-\epsilon_{i, j})^2}/N_k < \epsilon_{\rm tol}$ is satisfied, we solve for $f(k)$ with an initial guess $f_0(k) = [2\pi\,(1+{\rm e}^{\epsilon(k)})]^{-1}$ and the recursive relation
\begin{equation}
f_{j+1}(k) = f_0(k) + L(c_0, q) \circledast f_j(q),
\end{equation}
from which we get to evaluate (\ref{eq:linear_density_YY}). After watching all the unit conversions we get the linear density in particles per meter.

\end{document}